\documentclass[aps,prl,twocolumn,superscriptaddress]{revtex4}

\usepackage{graphicx}

\begin{document}

\preprint{}

\title{Uniform Mixing of High-$T_{\rm c}$ Superconductivity and Antiferromagnetism on a Single CuO$_2$ Plane in Hg-based Five-layered Cuprate}

\author{H. Mukuda}
\email[]{e-mail  address: mukuda@mp.es.osaka-u.ac.jp}
\author{M. Abe}
\author{Y. Araki}
\author{Y. Kitaoka }
\affiliation{Graduate School of Engineering Science, Osaka University, Toyonaka, Osaka 560-8531, Japan }

\author{K. Tokiwa}
\author{T. Watanabe}
\affiliation{Department of Applied Electronics, Science University of Tokyo, Noda, Chiba 278-8510, Japan}

\author{A. Iyo}
\author{H. Kito}
\author{Y. Tanaka}
\affiliation{National Institute of Advanced Industrial Science and Technology (AIST), Umezono, Tsukuba 305-8568, Japan}

\date{\today}

\begin{abstract}
We report a site selective Cu-NMR study on under-doped Hg-based five-layered high-$T_{\rm c}$ cuprate HgBa$_{2}$Ca$_{4}$Cu$_{5}$O$_{12+\delta}$ with a $T_{\rm c}$=72 K. 
Antiferromagnetism (AF) has been found to take place at $T_{\rm N}$=290 K, exhibiting a large antiferromagnetic moment of 0.67-0.69$\mu_{\rm B}$ at three inner planes (IP's). This value is comparable to the values reported for non-doped cuprates, suggesting that the IP may be in a nearly non-doped regime.
Most surprisingly, the AF order is also detected with $M_{\rm AF}$(OP)=0.1$\mu_{\rm B}$ even at two outer planes (OP's) that are responsible for the onset of superconductivity (SC). 
The high-$T_{\rm c}$ SC at $T_{\rm c}$ = 72 K can uniformly coexist on a microscopic level with the AF at OP's.
This is the first microscopic evidence for the uniform mixed phase of AF and SC on a single CuO$_2$ plane in a simple environment without any vortex lattice and/or stripe order. 
\end{abstract}


\maketitle

A remarkable aspect of high-$T_{\rm c}$ superconductivity (HTSC) is that there are many indications that their unique characteristics result from the competition between more than one type of order parameter. 
A possible coexistence of antiferromagnetism (AF) and HTSC is one of the remaining interesting problem in high-$T_{\rm c}$ cuprates. 
Although AF and HTSC are intimately related in high-$T_{\rm c}$ cuprates \cite{Anderson,Tanamoto,Kivelson,Zaanen,Fisher,Kampf,Zhang,Demler}, some HTSC compounds show AF/SC mixed phase, while others show microscopic separation between these two phases. 
Evidence for the AF/SC mixed phase exists in the excess oxygen doped La$_2$CuO$_{4+y}$ materials.  
Neutron scattering measurement detects the onset of the AF or spin-density-wave orders at the same temperature as the $T_{\rm c}$\cite{Lee}. 
On the other hand, microscopic probes such as STM revealed inhomogeneities of local electronic state and SC characteristics in nano-scale \cite{Pan}. 
Therefore, depending on material details, some HTSC compounds show AF/SC mixed phase, while others show microscopic separation between these two phases. 
Since such different physical effects can be obtained in materials that are so similar, it is still unclear if two phases coexist uniformly in these materials.

\begin{figure}[htbp]
\begin{center}
\includegraphics[width=1\linewidth]{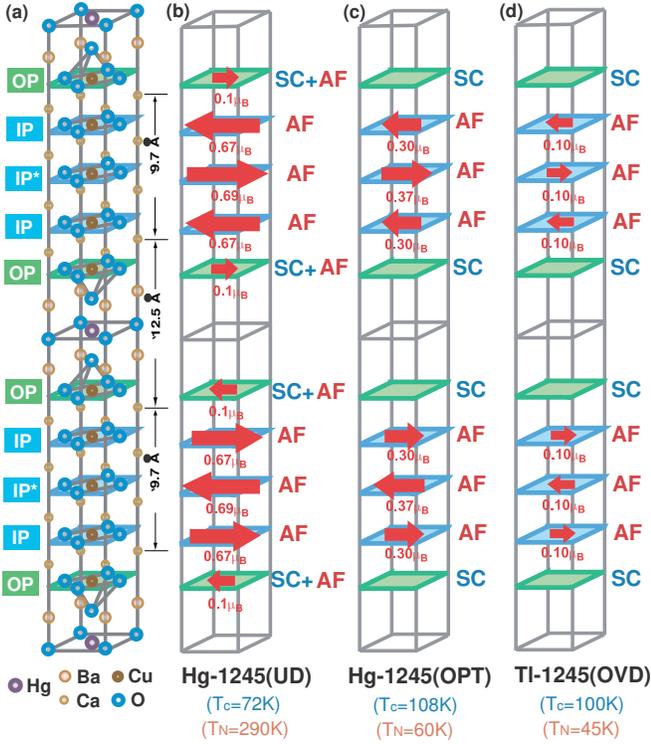}
\end{center}
\caption{(a): Crystal structure of Hg-based five-layered cuprates. Illustration of physical properties at each layer for (b) Hg-1245(UD), (c) Hg-1245(OPT)\cite{Kotegawa} and (d) Tl-1245(OVD). Here the AF moments($M_{\rm AF}$) are in the basal plane with an AF vector $(\pi/a,\pi/a)$\cite{MagneticStructure}. }
\label{fig:structures}
\end{figure}

Multilayered HTSC offers a naturally prepared heterostructure of AF planes and SC planes, which is a unique laboratory to investigate such underlying issue. 
Evidence for the coexistent phase of SC and AF in a unit cell has also been obtained recently in the five-layered HTSC HgBa$_2$Ca$_4$Cu$_5$O$_{12+\delta}$(Hg-1245(OPT)) by Kotegawa {\it et al} \cite{Kotegawa}(Fig.\ref{fig:structures}(c)). 
The series of multilayered HTSC include two types of CuO$_2$ planes, an outer CuO$_2$ plane (OP) in a pyramidal coordination and an inner plane (IP) in a square one with no apical oxygen, as shown in Fig.\ref{fig:structures}(a).  
A site selective Cu-NMR study enables us to unravel the layer-dependent electronic characters microscopically as follows \cite{Kotegawa,Kotegawa2001}; (1) The flat CuO$_2$ layers are homogeneously doped, ensured by the narrowest NMR line-width among the previous examples with very high quality to date, (2) Hole density distributes with a larger doping level at OP than at IP. 
Its charge imbalance increases as either a total carrier content and/or number of CuO$_2$ layers increase, (3) The study on Hg-1245(OPT) has revealed that the optimally doped two OP's are predominantly superconducting with $T_{\rm c}$ = 108 K, whereas the under-doped three IP's show the AF at $T_{\rm N}$ = 60 K, indicating the AF/SC mixed phase in a unit cell (see Fig.\ref{fig:structures}(c)).  
Theoretical approach to the coexistent phase of SC and AF was proposed recently by Mori and Maekawa that Josephson coupling between the SC planes through the AF plane leads to the coexistence of both phases with a rather high $T_{\rm c}$ value \cite{Mori}.
However, no mixing of SC and AF was reported in the heterostructures artificially grown by stacking integer numbers of layers of superconducting La$_{1.85}$Sr$_{0.15}$CuO$_4$ and antiferromagnetic insulator La$_2$CuO$_4$ \cite{Bozovic}. 
This type of AF/SC proximity effect was not observed in the artificially grown layer structures.

In this letter, we report systematic Cu-NMR studies on five-layered cuprates from under-doped Hg-1245(UD) to slightly overdoped Tl-1245(OVD), and compare with optimally-doped Hg-1245(OPT) in Ref.\cite{Kotegawa}. 

\begin{figure}[htbp]
\begin{center}
\includegraphics[width=0.7\linewidth]{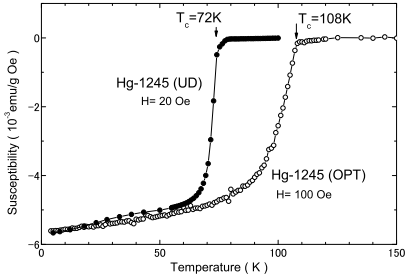}
\end{center}
\caption{DC susceptibility for Hg-1245(UD) after annealing and as-prepared Hg-1245(OPT)\cite{Kotegawa}. $T_{\rm c}$ is determined to be 72 K by a very sharp and single transition, indicating the successful removal of the excess oxygen uniformly from charge-reservoir. It was ensured by the similar superconducting volume fraction for both compounds.}
\label{fig:DCsusceptibility}
\end{figure}

Polycrystalline samples of Hg-1245 were prepared by the high-pressure synthesis technique as described elsewhere \cite{Tokiwa1996}. 
In order to reduce the carrier density, the as-prepared sample of Hg-1245 was annealed in Ar gas atmosphere for 280 hours. 
Figure \ref{fig:DCsusceptibility} shows the DC susceptibility measurement on Hg-1245(UD) after annealing, together with that of as-prepared one, optimally doped Hg-1245 (Hg-1245(OPT)) \cite{Kotegawa}. 
The $T_{\rm c}$ for the former is determined to be 72 K from a very sharp and single transition, indicating the successful removal of the excess oxygen uniformly from charge-reservoir layers. 
This evidences that the superconducting volume fraction is kept similar for both compounds. 
Figure \ref{fig:structures}(a) indicates the crystal structure of Hg-(Tl-)1245 including five CuO$_2$ layers (two pyramidal OP's and three square IP's). 
Note that IP$^*$ is the middle plane of three IP's. 

\begin{figure}[htbp]
\begin{center}
\includegraphics[width=0.8\linewidth]{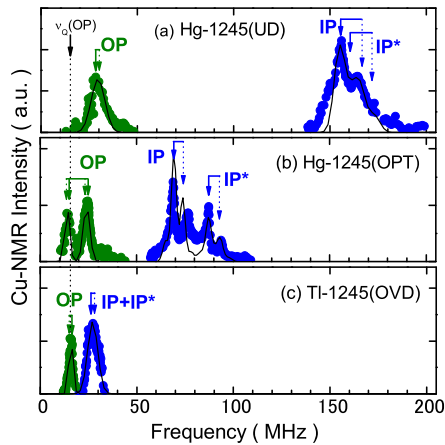}
\end{center}
\caption{Zero-field NMR spectra at 1.6 K for (a) Hg-1245(UD), (b) Hg-1245(OPT)\cite{Kotegawa} and (c) Tl-1245(OVD). The resonance spectrum from 140 to 180 MHz arises from IP's and IP$^*$, and the broad spectrum at 20$\sim$40MHz arises from OP. Solid and broken arrows show the signals from two isotopes $^{63}$Cu and $^{65}$Cu respectively. The solid line indicates the calculated spectrum using the parameters in this text.}
\label{fig:ZFNMR}
\end{figure}

The nuclear Hamiltonian is described in terms of the Zeeman interaction ${\cal H_Z}=-\gamma_{\rm N} \hbar {\bf H}\cdot{\bf I}$ and the nuclear quadrupole interaction ${\cal H_Q}=e^2qQ/4I(2I-1)(3I_z^2-I(I+1))$, where $\gamma_{\rm N}$ is the nuclear gyromagnetic ratio of Cu and $3e^2qQ/2I(2I-1)\equiv h\nu_Q$ is the nuclear quadrupole frequency. 
Figure \ref{fig:ZFNMR} shows the zero-field NMR spectrum at 1.6 K for Hg-1245(UD), together with that for Hg-1245(OPT) in the lower panel \cite{Kotegawa}. 
Two resonance peaks from 140 to 180 MHz arise from two isotopes $^{63}$Cu and $^{65}$Cu with slightly different nuclear gyromagnetic ratio $^{63,65}\gamma_{\rm N}$ at each site. 
This is the case for ${\cal H_Z} \gg {\cal H_Q}$. 
These spectra for Hg-1245(UD) allow to estimate the respective internal fields to be $H_{\rm int}$(IP)$=13.8$ T and $H_{\rm int}$(IP$^*)=14.2$ T at IP and IP$^*$ by fitting the spectra with the intensity ratio for two isotopes at IP and IP$^*$.  By using the hyperfine-coupling constant to be $H_{\rm hf}$(IP) = $-$20.7 T/$\mu_{\rm B}$ \cite{estimation}, the AF ordered moments $M_{\rm AF}$ at IP and IP$^*$ are estimated to be 0.67 and 0.69 $\mu_{\rm B}$, respectively. 
Note that these values are almost comparable to the respective $M_{\rm AF}$=0.50 and 0.64 $\mu_{\rm B}$ in the non-doped La$_2$CuO$_4$ \cite{Vaknin} and YBa$_2$Cu$_3$O$_6$ \cite{Rossat-Mignod}. 
From the $\mu$SR measurement, the AF has been found to take place at $T_{\rm N}=290$ K \cite{TokiwaUP} that is also comparable to $T_{\rm N}=325$ K and 415 K reported in the non-doped La$_2$CuO$_4$ \cite{Vaknin} and YBa$_2$Cu$_3$O$_6$ \cite{Rossat-Mignod}, respectively. 
These values are larger than $M_{\rm AF}$(IP)=0.30 $\mu_{\rm B}$, $M_{\rm AF}$(IP$^*$)=0.37 $\mu_{\rm B}$ and $T_{\rm N}$= 60 K for Hg-1245(OPT) which is in the metallic AF regime \cite{Kotegawa}. 
It is hence likely for the IP and the IP$^*$ in Hg-1245(UD) to be in a non-doped AF regime. 
On the other hand, from the spectra in Fig.\ref{fig:ZFNMR}(c), $M_{\rm AF}$(IP's)=0.1$\mu_{\rm B}$ is estimated for Tl-1245(OVD) with $T_{\rm N}\sim$45K\cite{Kotegawa}, suggesting that the hole density at IP's is located near a critical value above which AF collapses as discussed later. 

Most surprisingly, the zero-field NMR spectrum at OP for Hg-1245(UD) was observed over a broad frequency range 20$\sim$40 MHz that is about twice larger than the Cu-NQR frequency ($\nu_{\rm Q}$(OP)=16 MHz) observed for the case of paramagnetic OP in Tl-1245(OVD) and Hg-1245(OPT)(shown by the broken line in Fig.\ref{fig:ZFNMR}).
The observed broad spectrum was reproduced by an internal field of about $H_{\rm int}$(OP)=2.4 T with a fixed parameter of $\nu_Q$(OP)=16 MHz\cite{estimation}, as indicated in Fig.\ref{fig:ZFNMR}. 
This internal field is larger than not only the calculated dipole field($\sim$70 Oe) but also 0.54 T for the case of proximity effect at the OP from AF ordered IP\cite{Kotegawa}.
A size of ordered moment $M_{\rm AF}$(OP)= 0.1$\mu_{\rm B}$ is estimated using the relation of $H_{\rm hf}$(OP)= $-$26 T/$\mu_{\rm B}$ for the OP for Hg-1245(UD) \cite{estimation}. 
The distribution of $M_{\rm AF}$ is illustrated in Fig.\ref{fig:structures}. 
The $H_{\rm int}$(OP)=2.4 T at OP's is about five times larger than $H_{\rm int}$(OP)=0.5 T for Hg-1245(OPT), while $M_{\rm AF}$(IP)$\sim 0.67\mu_{\rm B}$ is about twice larger than $M_{\rm AF}$(IP)$\sim 0.3\mu_{\rm B}$ for Hg-1245(OPT).  
Therefore, it is concluded that $H_{\rm int}$(OP)=2.4 T at the OP's for Hg-1245(UD) is not only transferred from $M_{\rm AF}$(IP), but also AF order manifests itself in the OP at low temperatures. When noting that no NMR signal was observed in the vicinity of $\nu_Q$(OP)=16 MHz that would be expected for the paramagnetic OP, a possibility of phase separation into paramagnetic SC domains and AF domains is experimentally excluded. Instead, all of the Cu sites at OP are microscopically demonstrated to undergo a uniform AF order with about $M_{\rm AF}$(OP)$\sim 0.1\mu_{\rm B}$. 
This is the first microscopic evidence for the uniform mixed phase of AF and SC on a single CuO$_2$ plane (OP) in a simple environment without any vortex lattice \cite{Lake} and/or stripe order \cite{Tranquada}. 

We consider that a nearly perfect flatness of CuO$_2$ planes is a key for the first observation of uniform mixed state of AF and SC in the cuprates. 
In mono- and bi-layered cuprates, the phase separation has been found by microscopic probes due to inhomogeneity of the electronic state at CuO$_2$ plane \cite{Pan,Ishida}, because the buckling of CuO$_2$ planes may be inevitable owing to doping holes for LSCO, for instance.
On the other hand, the CuO$_2$ layers in the multilayered cuprates are not directly affected by some disorder effect introduced into the charge reservoir layers through the carrier doping process. 
The flatness is guaranteed by the narrowest Cu-NMR line-width about 50 Oe (150 Oe) for the IP (OP) in the polycrystalline of Hg-based systems \cite{Kotegawa}. 

\begin{figure}[htbp]
\begin{center}
\includegraphics[width=0.8\linewidth]{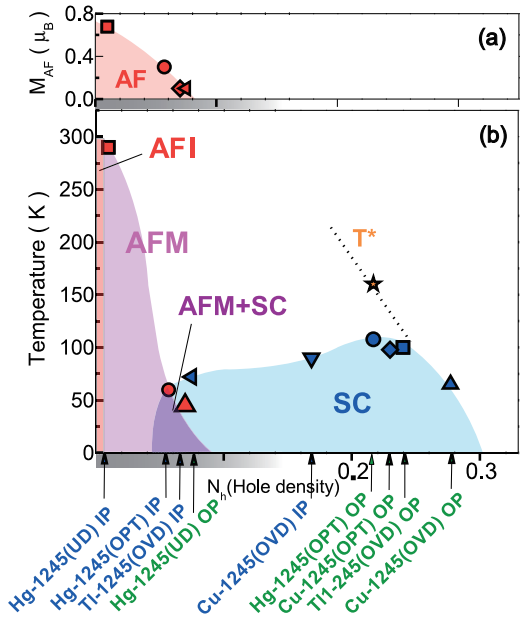}
\end{center}
\caption{(b) A phase diagram derived from the physical properties at the IP and the OP for various five-layered HTSC. Here we denote AF insulator and metal phase as AFI and AFM. $T^*$ is a pseudogap temperature deduced from a decrease of $1/T_1T$ at the OP for Hg-1245(OPT). The panel (a) shows the variation in size of AF ordered moment $M_{\rm AF}$ as the function of hole density. Since the hole density of underdoped region indicated by gray line could not be determined precisely, we plot by using the relative hole density as described in text\cite{holedensity}.}
\label{fig:Phasediagram}
\end{figure}

We here address a novel phase diagram in Fig. \ref{fig:Phasediagram}(b) that is derived from the systematic Cu-NMR studies on the Hg-, Tl- and Cu-based five-layered HTSC via using the local hole density at OP and IP\cite{holedensity}.
This phase diagram differs from the global phase diagram of the HTSC reported so far, for instance, such as LSCO. It should be noted that the nearly non-doped AF in IP and IP$^*$ takes place, whereas inhomogeneous magnetic phases such as spin-glass phase or stripe phase are not observed at both IP's and OP's. Instead, the presence of the doped AF metallic (AFM) phase at IP and IP$^*$ is remarkable at the boundary between AF insulating (AFI) phase and SC \cite{Kotegawa}. This differs from the case of LSCO where the disorder-driven  magnetic phases exist between the AFI phase in $N_h<$ 0.02 and the SC phase in $N_h>$ 0.05.  It is noted that the AF for IP is extended to a higher hole density due to the flatness of CuO$_2$ plane with no apical oxygen and the homogeneous distribution of carrier density. By contrast, the prototype phase diagrams reported thus far are under the inevitable disorder effect associated with the chemical substitution introduced into the CuO$_2$ out-of-planes.
Here the hole densities in the underdoped region ($N_h<$0.13) are determined  by using the relative hole density suggested from Knight shift and $T_1$ measurements. For example, $N_h$ at the IP's for Tl-1245(OVD) is slightly larger than that for Hg-1245(OPT). 
As for the OP in Hg-1245(UD), $N_h$ may be compatible with that at the IP's of Tl-1245(OVD) because the size of $M_{\rm AF}$ is the same for both.
It is this global phase diagram to make us convince the presence of the AF/SC uniformly mixed phase at OP of Hg-1245(UD), where $M_{\rm AF}$=0.1$\mu_{\rm B}$ and $T_c$=72 K. 
The OP of Hg-1245(UD) and IP's of Tl-1245(OVD) might be located just on a region where AF and SC confront with each other.

Unexpectedly in Hg-1245(UD), it has been revealed that the nearly non-doped AF IP's and the AF/SC uniformly mixed OP's are alternatively stacked as (AF+HTSC)/AF/(AF+HTSC) along the $c$-axis, as shown in Fig.\ref{fig:structures}(b). 
The $T_{\rm N}$ for AF may be determined through the interlayer magnetic coupling. 
Although the AF order for IP's occurs below $T_N=290$ K, $T_{\rm N}$ inherent to the OP is not known because the $T_{\rm N}$ of this compound is dominated by the interlayer magnetic coupling between the nearly non-doped AF IP's and the AF/SC mixed OP's. Surprisingly this interlayer magnetic coupling does not prevent the onset of SC with the high value of $T_{\rm c}$=72 K, even though the SC uniformly coexisting with the AF is significantly separated over more than 10\AA\  by three nearly non-doped AF IP's below $T_{\rm N}$ = 290 K.  It remains a mystery why it keeps a high-$T_{\rm c}$ value. 
According to the theoretical studies on multilayered cuprates \cite{Chakravarty,Mori}, the $T_{\rm c}$ for SC may be determined through the quantum tunneling of Cooper pairs between the layers.  The fact that HTSC can keep a high value of $T_{\rm c}$ even in the mixed state with AF on the same CuO$_2$ plane suggests that HTSC and AF phases are nearly degenerate and that the mechanism of SC is magnetic in origin.  An explanation for the uniform mixing of AF and SC phases at OP is relevant with the theoretical prediction based on the SO(5) model \cite{Zhang,Demler}, that is, the correlation length for the superconducting proximity effect across an AF should be very long, and hence supercurrent should flow through the naturally prepared thick heterostructure of (AF+HTSC)/ AF /(AF+HTSC) in the present multilayered systems.  Thus as long as HTSC and AF phases are nearly degenerate, the proximity effect ought to be strong.  

In conclusion, we have found that the under-doped Hg-1245 has the nearly non-doped three IP's with $T_{\rm N}$ = 290 K and $M_{\rm AF}$(IP)$\approx$0.67-0.69$\mu_{\rm B}$ and the superconducting two OP's with $T_{\rm c}$ = 72 K. The AF order was also detected with $M_{\rm AF}$(OP)$=$0.1$\mu_{\rm B}$ even at the OP that are responsible for the onset of SC. This finding  provides the first microscopic evidence for the uniform mixing of AF and SC at a single CuO$_2$ plane in  HTSC.  Although the AF/SC mixed CuO$_2$ planes are significantly separated by three non-doped AF layers, the onset of AF does not prevent the occurrence of SC with the high value of $T_{\rm c}=72$ K.  In a large class of materials including HTSC, organic and heavy-fermion superconductors, AF and SC occur in close proximity to each other. A genuine uniform mixed phase of AF and SC has been observed in the pressure versus temperature phase diagrams in several heavy-fermion systems \cite{Kitaoka,YKawasaki,SKawasaki}. The present results give a hint to gain insight into a mechanism in strongly correlated superconductivity in general. 

The authors would like to thank H. Kotegawa, S. Maekawa, M. Mori, K. Ishida and G.-q. Zheng for their valuable discussions and comments, and M. Yashima, S. Kawasaki for their experimental supports.  This work was supported by Grant-in-Aid for Creative Scientific Research (15GS0213) from the Ministry of Education, Culture, Sports, Science and Technology (MEXT) and the 21st Century COE Program (G18) by Japan Society of the Promotion of Science (JSPS).


\clearpage


\begin{references}
\bibitem{MagneticStructure} The magnetic structure is speculated to be antiparallel between the adjacent layers. If the $M_{\rm AF}$'s in the plane were ferromagnetically aligned between the layers, the internal field at each layer should be affected not only by the on-site $M_{\rm AF}$ but also by the transferred hyperfine fields with negative sign against the on-site one that are induced by $M_{\rm AF}$'s at the adjacent layers. In this case, the on-site $M_{\rm AF}$ is roughly calculated as 0.7 $\sim$ 0.8 $\mu_{\rm B}$ for the IP$^*$ of Hg-1245(UD), that amounts to an unreasonably large value.
\bibitem{estimation} $\nu_Q$(OP), $H_{\rm hf}$(IP) and $H_{\rm hf}$(OP) are assumed to be the same with the values of Hg-1245(OPT), because it is not so different for the same Hg-based five-layered system.
\end{references}
\end{document}